\documentclass[aps,prl,reprint,superscriptaddress,nofootinbib,longbibliography,floatfix]{revtex4-2}

\usepackage{amsmath,amssymb}
\usepackage{graphicx}
\usepackage{bm}
\graphicspath{{figures/}}
\usepackage[colorlinks=true,allcolors=blue]{hyperref}

\begin{document}

\title{Measurement and reload costs in direct quantum simulation of nonlinear waves}

\author{Ziqing Guo}
\email{ziqing.guo@ttu.edu}
\affiliation{Texas Tech University, Lubbock, TX 79409, USA}

\author{Viraj Dsouza}
\email{virajdanieldsouza@gmail.com}
\affiliation{BosonQ Psi Corp, 235 Harrison St, Syracuse, NY 13202 USA}

\author{Alex Khan}
\affiliation{BosonQ Psi Corp, 235 Harrison St, Syracuse, NY 13202 USA}

\author{Abhishek Chopra}
\email{abhishekchopra@bqpsim.com}
\affiliation{BosonQ Psi Corp, 235 Harrison St, Syracuse, NY 13202 USA}

\author{Rut Lineswala}
\email{rutlineswala@bqpsim.com}
\affiliation{BosonQ Psi Corp, 235 Harrison St, Syracuse, NY 13202 USA}

\author{Ziwen Pan}
\email{ziwen.pan@ttu.edu}
\affiliation{Texas Tech University, Lubbock, TX 79409, USA}

\date{July 7, 2026}

\begin{abstract}
Quantum processors encode an $N$-point field in $\log_2 N$ qubits, which renders
nonlinear wave equations an important application for quantum simulation. Nonlinear
evolution, however, requires the field values themselves, and these are not
directly accessible without quantum measurement. Existing algorithms circumvent
this measurement through linear embeddings and state copies, thereby obscuring its
cost within the truncation order, the auxiliary dimensions, and the state
preparation. In order to expose this cost, a hybrid split-step solver is proposed
in which the field is measured, updated classically, and reloaded at every step,
with all shots and gates accounted for in a single cost-and-error model. Since the
entire field is available at every step, a property unavailable to linear
approximations in strongly nonlinear regimes, the design of the solver reduces to a
budgeting problem over the timestep, the polynomial degree, and the shot count. The
coherent kernels of the solver are validated on superconducting hardware. An
identical structure and bottleneck govern the viscous Burgers' equation in one and
two dimensions. Because every step reads the full field, the
quantum cost per step, measured as circuit depth multiplied by measurement shots,
exceeds the classical cost with increasing grid size. The framework consequently
identifies a coherent, measurement-free nonlinear update as the quantitative
target that any end-to-end advantage must meet.
\end{abstract}

\maketitle

\section{Introduction}
\label{sec:intro}

The numerical solution of a nonlinear partial differential equation (PDE) involves
representing the field on a grid consisting of $N$ points and advancing it
temporally \cite{strang1968}. In a classical computing framework, one field value
is stored per grid point, resulting in memory usage that increases linearly with
$N$. Furthermore, $N$ increases exponentially with the spatial dimension when
resolution is held constant. Conversely, a quantum register encodes the entire
field within the amplitudes of $n=\log_2 N$ qubits, achieving an exponential
reduction in memory requirements \cite{lloyd1996,harrow2009}. The linear component
of the dynamics is evolved with a circuit depth that is polynomial in $n$, due to
its diagonal nature in Fourier space \cite{jin2024bull}. However, the nonlinear
term is non-unitary and is contingent upon the field value at each point, which the
register maintains as an amplitude rather than a directly accessible number. The
cubic nonlinear Schr\"odinger equation (NLSE) for the complex field $\psi(x,t)$
\cite{agrawal2019,pitaevskii2003},
\begin{equation}
i\,\partial_t\psi = -\tfrac{1}{2}\,\partial_x^2\psi - \kappa\,|\psi|^2\psi,
\label{eq:nlse}
\end{equation}
in which $\kappa$ denotes the nonlinearity strength, separates these two components
cleanly. Its linear part is exactly diagonal in Fourier space, and its
nonlinearity constitutes an exactly diagonal phase in
position space determined by the local intensity $|\psi_j|^2$. Each factor of a
split-step update is therefore exact, and the sole remaining obstruction is that
$|\psi_j|^2$ cannot be read from the register without measurement.

Several strategies have been developed to accommodate this nonlinearity.
Nonlinear-feedback schemes \cite{lloyd2020,leyton2008} apply the nonlinearity
directly to the state, although every step consumes several fresh copies of that
state. Carleman and lattice-Boltzmann linearizations
\cite{liu2021,krovi2023,wu2025,budinski2021,sanavio2024} restore unitary evolution
by lifting the dynamics to a truncated ladder of field moments, the dimension of
which grows with the truncation level and the convergence of which requires weak
nonlinearity or dissipation. Koopman--von Neumann embeddings \cite{joseph2020} and
Schr\"odingerization map dissipative or non-unitary dynamics onto Hamiltonian
evolution in an enlarged space \cite{jin2024prl,jin2023pra}. In particular, hybrid
schemes divide the computation between processors: the quantum processor solves
linear subproblems while nonlinear variables are advanced classically
\cite{jin2024bull,jin2026} or a parametrized wavefunction is fit variationally to
the evolved state \cite{lubasch2020}. In each case, the reported resource count is
expressed in terms of the truncation order, the dimension of the auxiliary space,
or the cost of state preparation.

\begin{figure}[tbp]
\centering
\includegraphics[width=\columnwidth]{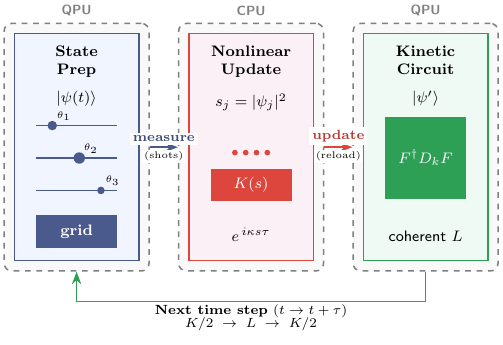}
\caption{Diagram of the hybrid solver implementing the nonlinear half-kicks
$K_{\Delta t/2}$ as a measure--update--reload cycle between QPU and CPU, using
computational-basis sampling of the intensities $s_j=|\psi_j|^2$, with the linear
propagator $L_{\Delta t}=F^\dagger D_k F$ applied coherently in the spectral
basis, for one split step $K_{\Delta t/2}\,L_{\Delta t}\,K_{\Delta t/2}$ advancing
$t\to t+\Delta t$.}
\label{fig:hybrid}
\end{figure}

In this work, the classical split-step factorization
\cite{strang1968,agrawal2019,taha1984}
\begin{equation}
\psi(t+\Delta t) \approx K_{\Delta t/2}\, L_{\Delta t}\, K_{\Delta t/2}\,\psi(t),
\label{eq:splitstep}
\end{equation}
is applied directly to the amplitude-encoded field over a timestep $\Delta t$. The
nonlinear update $K_\tau$ is implemented as a measure--update--reload cycle
alongside the coherent linear propagator $L_{\Delta t}$ (Fig.~\ref{fig:hybrid}),
with arithmetic based on a vectorized encoding \cite{guo2025}. Shot
count, reload latency, and gate depth thereby enter a single cost model, which
reduces solver design to the allocation of a fixed budget across timestep,
polynomial degree, and shot count. The circuits are detailed in
Sec.~\ref{sec:setup} and the cost model in Sec.~\ref{sec:cost}.

The nonlinear phase accumulated at a grid point during one step,
$\kappa|\psi_j|^2\Delta t$, determines the operating regime of the solver.
Low-order truncations of the kick fail once this phase approaches unity, which the
controller counters by reallocating a fixed measurement budget across timestep,
degree, and shots. We also verify the coherent kernels of the cycle on an IBM
Heron-architecture machine, and recover the same sampling floor for the viscous
Burgers' equation in one and two dimensions. Because
every step reads the whole field, the per-step quantum cost grows more rapidly with
grid size than that of the classical split step, and the pipeline therefore retains
no end-to-end advantage at fixed precision. A coherent, measurement-free nonlinear
update is the decisive requirement, and the per-step scaling supporting this
conclusion is derived in Sec.~\ref{sec:resource}.

\section{Algorithm setup}
\label{sec:setup}

The field is represented by the amplitude-encoded grid state on $N=2^n$ points of
spacing $\Delta x$,
\begin{equation}
|\psi\rangle = \sum_{j=0}^{N-1}\psi_j\,|j\rangle,
\qquad
\sum_j |\psi_j|^2 = 1,
\label{eq:state}
\end{equation}
in which $|j\rangle$ denotes the computational basis state whose $n$-bit binary
label is the grid index $j$, and $\psi_j$ the field value sampled at that grid
point. One split step of Eq.~\eqref{eq:splitstep} is applied to this state per
measure--update--reload cycle of Fig.~\ref{fig:hybrid}, with $K_\tau$ the nonlinear
half-kick over a time $\tau=\Delta t/2$ and $L_{\Delta t}$ the linear propagator
over $\Delta t$. The symmetric ordering $K\,L\,K$ of Eq.~\eqref{eq:splitstep} is
the Strang form, whose local error is $\mathcal{O}(\Delta t^3)$. The method
comprises three kernels, each of which carries a latency and an infidelity in the
cost model of Sec.~\ref{sec:cost}.

\begin{figure}[tbp]
\centering
\includegraphics[width=\columnwidth]{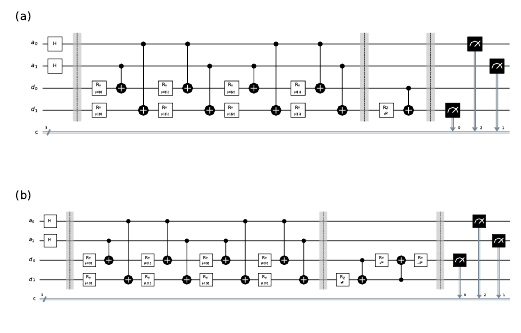}
\caption{Vectorized arithmetic circuits of \textbf{PolyKick}, shown for two
address qubits ($a_0,a_1$) and two data qubits ($d_0,d_1$). (a)~For the
elementwise product, uniformly controlled $R_y$ rotations load both operands in
parallel, and the $R_z(\pi/2)$-CX block writes the product onto the measured
channel. (b)~For the weighted sum, a controlled-$R_y$ recombination realizes
$(1-w)\,x + w\,y$. The
loading depth of $\Theta(N)$ controlled-NOT (CX) gates sets the reload latency
$\tau_{\rm rel}(N)$ of Eq.~\eqref{eq:reload}.}
\label{fig:polykick}
\end{figure}

\textbf{StateReload} is the state-preparation boundary, which returns a classical
vector $v\in\mathbb{C}^{N}$ to the $n$-qubit register initialized in
$|0^n\rangle$ as
\begin{equation}
R(v)\,|0^n\rangle = \frac{1}{\|v\|_2}\sum_{j=0}^{N-1} v_j\,|j\rangle,
\label{eq:reload}
\end{equation}
with latency $\tau_{\rm rel}(N)$ and infidelity $\eta_{\rm rel}$. General state
preparation admits numerous constructions \cite{grover2002,araujo2021}. We give
the vectorized loader \cite{balewski2024}, for which a $\Theta(N)$ CX depth at a
two-qubit gate time of $50$~ns yields $\tau_{\rm rel}\sim 1.6\,\mu$s at $N=32$,
comparable to the coherent linear step itself. \textbf{SpecProp} is the unitary
spectral step
\begin{equation}
L_{\Delta t} = F_N^\dagger\!\left(\sum_{m=0}^{N-1}
e^{-ik_m^2\Delta t/2}\,|m\rangle\langle m|\right)F_N,
\label{eq:specprop}
\end{equation}
and we note that the classically easy simulation framework \cite{javadi2024}
enables this step to be validated against the exact statevector, where $F_N$
denotes the quantum Fourier transform (QFT) on the $n$ address qubits and the
bracketed diagonal operator, written $D_k$ in Fig.~\ref{fig:hybrid}, applies the
free-dispersion phase $e^{-ik_m^2\Delta t/2}$ to spectral mode $m$. The discrete
wavenumbers of the grid are $k_m=2\pi m/(N\Delta x)$ for $m<N/2$ and
$k_m=2\pi(m-N)/(N\Delta x)$ otherwise. Its coherent time
$T_{\rm lin}(n,\Delta t)$ is the
$n(n-1)/2$ two-qubit gates of an exact QFT \cite{coppersmith1994} together with
the single-qubit rotations of the diagonal phase at precision $b_\phi$, evaluated
at calibrated gate times. \textbf{PolyKick} approximates
the exact half-kick $(K_\tau\psi)_j = \psi_j\,e^{i\kappa|\psi_j|^2\tau}$ by
truncating the exponential series at degree $d$,
\begin{equation}
P_d(i\theta_j) = \sum_{\ell=0}^{d}\frac{(i\theta_j)^\ell}{\ell!},
\qquad
\theta_j = \kappa\tau\,\widehat{p}_j,
\label{eq:polykick}
\end{equation}
followed by renormalization,
\begin{equation}
\psi_j^{+} = \frac{\psi_j\,P_d(i\theta_j)}{\|\psi\,P_d\|_2},
\qquad
\eta_{\rm poly} = \bigl|\,\|\psi\,P_d\|_2 - \|\psi\|_2\,\bigr|,
\label{eq:renorm}
\end{equation}
in which $\widehat{p}_j$ denotes the estimate of the exact intensity $|\psi_j|^2$
obtained from computational-basis samples, $\theta_j$ the resulting nonlinear phase
at site $j$, so that $P_d(i\theta_j)\to e^{i\theta_j}$ as $d\to\infty$, and
$\psi\,P_d$ the vector with entries $\psi_j P_d(i\theta_j)$. The norm defect
$\eta_{\rm poly}$ measures the departure of the truncated factor from a pure
phase, which is unity in modulus. The vector
$\psi^{+}$ is returned to \textbf{StateReload}. The product and weighted-sum
circuits realizing Eq.~\eqref{eq:polykick}, constructed from the QCrank vectorized
encoding \cite{balewski2024} and the EHands arithmetic primitives
\cite{balewski2025}, are shown in Fig.~\ref{fig:polykick}; a tolerance-matched
quantum singular value transformation (QSVT) comparison is provided in
Sec.~\ref{app:qsvt}; and splitting-order and dispersion ablations are collected in
Sec.~\ref{app:ablation}. Should reversible arithmetic implement this kick without
measurement, Eq.~\eqref{eq:renorm} would be replaced by that function and the
sampling term would be removed from the cost model.

\begin{figure}[tbp]
\centering
\includegraphics[width=\columnwidth]{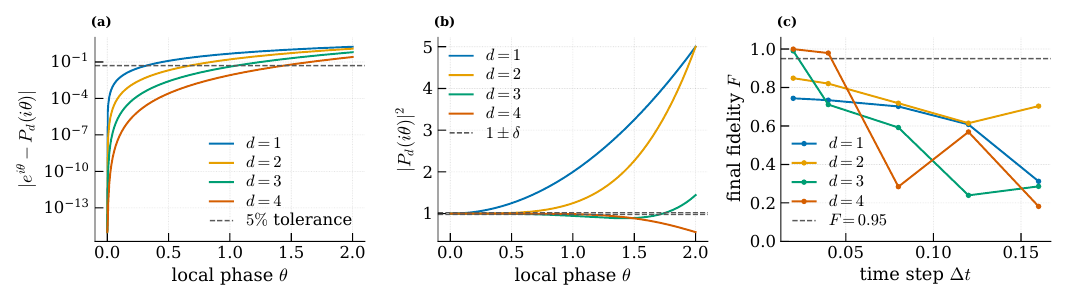}
\caption{Accuracy diagnostics of the Taylor half-kick $P_d(i\theta)$ of
Eq.~\eqref{eq:polykick}, where $d$ is the polynomial truncation degree and
$\theta=\kappa|\psi_j|^2\Delta t$ is the nonlinear phase accumulated at one grid
point per timestep $\Delta t$ at nonlinearity strength $\kappa$. (a)~Phase-factor
error $|e^{i\theta}-P_d(i\theta)|$ for $d=1$--$4$; the dashed line marks a
representative $5\%$ error tolerance. (b)~Squared norm $|P_d(i\theta)|^2$ of the
truncated phase factor; dashed lines mark $1\pm\delta$ at the controller
norm-defect threshold $\delta=0.02$. (c)~Overlap fidelity
$F=|\langle\psi_{\rm ref},\psi\rangle|^2$ between the normalized final states of
the hybrid solver and a fine-step classical split-step Fourier reference, versus
$\Delta t$, for a soliton of initial amplitude $A=1.8$ evolved on $N=32$ grid
points at $\kappa=50$ to final time $T=0.48$; the dashed line marks $F=0.95$.}
\label{fig:accuracy}
\end{figure}

The accuracy limits of this construction are presented in Fig.~\ref{fig:accuracy}.
The error of the truncated kick is governed by the Taylor remainder of
Eq.~\eqref{eq:polykick}, bounded by $\theta^{d+1}/(d+1)!$, so that each degree
remains accurate only while the local phase $\theta$ is small, as illustrated in
Fig.~\ref{fig:accuracy}(a). Once $\theta$ approaches unity, the remainder terms
cease to decay with order and the truncated factor is no longer a pure phase. The
norm defect $\eta_{\rm poly}$ of Eq.~\eqref{eq:renorm} grows with a higher power of
$\theta$, as demonstrated in Fig.~\ref{fig:accuracy}(b) by the separation between
low and high degrees, and it violates tolerance before the phase error does. At a
sufficiently small timestep, $\theta = \kappa|\psi_j|^2\tau$ is returned below both
thresholds and the final fidelity is restored, as shown in
Fig.~\ref{fig:accuracy}(c), which reflects the fact that degree and timestep
control the same variable $\theta$. Truncation error, norm defect, and timestep
therefore constitute the three levers adapted by the controller of
Sec.~\ref{sec:cost}.

\section{Cost and error model}
\label{sec:cost}

This section establishes how gate time, measurement shots, and approximation error
combine within a single step. The method is initialized with a timestep
$\Delta t_0$, a polynomial degree $d\in\{2,4,6\}$, and a shot count $S$ per
half-kick. In order to render each cost explicit, the kernels of
Sec.~\ref{sec:setup} are collected into a single latency-and-error model that is
adjusted by the controller at every step.

One Strang step contains two nonlinear half-kicks, two reloads at the
nonlinear--classical boundary, and one coherent spectral propagation. A proposed
step is accepted when its norm defect satisfies $\eta_{\rm poly}\le\delta$ and is
rejected otherwise; a rejected attempt is retried with the timestep halved. For a
step employing degree $d$, $S$ shots per half-kick, and timestep $\Delta t$, the
latency is modeled as
\begin{equation}
\begin{split}
T_{\rm step} = {}& 2\,T_{\rm meas}(S,N) + 2\,\tau_{\rm rel}(N) \\
&+ T_{\rm lin}(n,\Delta t) + T_{\rm class}(N,d),
\end{split}
\label{eq:tstep}
\end{equation}
in which the factors of two count the two half-kicks of one Strang step. Every
shot consumes the state, so each half-kick re-prepares it $S$ times and the
sampling time is $T_{\rm meas}(S,N)=S\,[\tau_{\rm rel}(N)+\tau_{\rm ro}]$, with
$\tau_{\rm ro}$ the readout time. The remaining terms are the single reload of the
updated vector $\psi^{+}$ of Eq.~\eqref{eq:renorm} at latency $\tau_{\rm rel}(N)$,
the coherent spectral time $T_{\rm lin}(n,\Delta t)$ of Eq.~\eqref{eq:specprop},
and the classical polynomial update $T_{\rm class}(N,d)=\mathcal{O}(Nd)$. The
factor $S$ multiplying the $\Theta(N)$-depth reload is the origin of the per-step
work of Sec.~\ref{sec:resource}. The total modeled runtime sums
Eq.~\eqref{eq:tstep} over accepted and rejected attempts, and reports modeled
algorithmic runtime rather than local wall-clock time.

\subsection{Sampling requirement}

In prior hybrid schemes, field information is estimated mode by mode through
Hadamard tests \cite{weng2026} or absorbed into variational state-preparation
assumptions \cite{lubasch2020}. We give a single multinomial experiment per
half-kick. Drawing $S$ computational-basis shots from $|\psi\rangle$ yields bin
counts $c_j$ that are multinomially distributed, and the estimator
$\widehat{p}_j=c_j/S$ recovers all $N$ intensities from the same $S$ shots, with
variance
\begin{equation}
\mathrm{Var}(\widehat{p}_j) = \frac{p_j(1-p_j)}{S},
\label{eq:var}
\end{equation}
where $p_j=|\psi_j|^2$ is the exact intensity at site $j$.

The shot count follows from the phase tolerance in two steps. First,
$\theta_j=\kappa\tau\,\widehat{p}_j$ is linear in the estimate, so the induced
phase error is $\delta\theta_j=\kappa\tau(\widehat{p}_j-p_j)$ and an intensity error
$|\widehat{p}_j-p_j|\le\varepsilon_p$ gives
\begin{equation}
|\delta\theta_j| \le \kappa\tau\,\varepsilon_p,
\label{eq:phaseerr}
\end{equation}
so a target phase tolerance $\beta_{\rm shot}$ requires
$\varepsilon_p\le\beta_{\rm shot}/\kappa\tau$. Second, Hoeffding's inequality
bounds the failure probability of one bin by $2e^{-2S\varepsilon_p^2}$, and a union
bound over the $N$ bins imposes $2Ne^{-2S\varepsilon_p^2}\le\alpha$, which inverts
to the sufficient shot count
\begin{equation}
S \ge \frac{1}{2\varepsilon_p^2}\,\log\frac{2N}{\alpha},
\label{eq:hoeffding}
\end{equation}
with $\alpha$ the probability that any of the $N$ estimates exceeds
$\varepsilon_p$. The controller below employs Eq.~\eqref{eq:hoeffding} to
determine $S$ from the target $\beta_{\rm shot}$. The
resulting per-step shot count at the benchmark operating point of
Sec.~\ref{sec:nlse} is reported in Fig.~\ref{fig:shots} of
Sec.~\ref{app:diagnostics}, where it already exceeds the per-step budget of a
near-term device.

Eq.~\eqref{eq:hoeffding} bounds each site separately, whereas the reload of
Eq.~\eqref{eq:reload} consumes the entire profile at once, so that the relevant
accuracy is aggregate rather than per-site. Recovery of an $N$-outcome distribution
to total-variation error $\varepsilon$ requires $\Theta(N/\varepsilon^2)$ samples
\cite{canonne2020}, and no measurement strategy applied to copies of
$|\psi\rangle$ circumvents this count, independent of the circuit used to prepare
or process the state. This lower bound constitutes the floor invoked in
Sec.~\ref{sec:resource}. The controller introduced below does not remove it, but
reallocates a fixed budget across timestep, degree, and shots, as quantified in
Sec.~\ref{sec:nlse}.

\subsection{Deterministic and stochastic errors}

The local error $e_{\rm loc}$ of a single accepted step is bounded by five
contributions,
\begin{equation}
\begin{split}
e_{\rm loc} \le {}& C_S\,\Delta t^3
+ \max_j\frac{|\kappa\tau p_j|^{d+1}}{(d+1)!} \\
&+ \kappa\tau\,\max_j|\widehat{p}_j-p_j|
+ \eta_{\rm rel} + \eta_{\rm poly},
\end{split}
\label{eq:eloc}
\end{equation}
in which $C_S$ is the Strang local-error constant for the smooth solution class
under consideration. The terms are, respectively, the splitting error, the Taylor
truncation error of Eq.~\eqref{eq:polykick}, the sampling-induced phase noise of
Eq.~\eqref{eq:phaseerr}, the reload infidelity of Eq.~\eqref{eq:reload}, and the
norm defect of Eq.~\eqref{eq:renorm} before renormalization. Over $M$ accepted
steps reaching final time $T$, a conservative stability statement is
\begin{equation}
e(T) \lesssim e^{LT}\sum_{m=1}^{M} e_{\rm loc}^{(m)},
\label{eq:stability}
\end{equation}
in which $e(T)$ is the accumulated error at time $T$, $e_{\rm loc}^{(m)}$ the
local error of step $m$, and $L$ a stability constant. Two contributions enter
$L$: the continuous NLSE flow, bounded on the soliton manifold by standard a
priori energy estimates, and the reload map, which is not bounded in closed form
here. The controller rejection rate serves as an operational proxy for $L$, since
a rising rate signals that $L\Delta t$ approaches order unity, which the
timestep-shrink response mitigates. Eq.~\eqref{eq:eloc} is therefore used as a
controller model rather than a sharp bound on global fidelity under hardware
noise.

\subsection{Controller}

For each proposed step, the controller selects the lowest degree $d\in\{2,4,6\}$
satisfying the truncation tolerance, determines $S$ from the target
$\beta_{\rm shot}$, and rejects the attempt if the norm defect $\eta_{\rm poly}$
exceeds $\delta$. A rejected attempt reduces $\Delta t$ by a factor $\gamma=0.5$,
subject to the lower bound $\Delta t_{\min}=0.01$, whereas sustained accepted steps
permit $\Delta t$ to increase toward its initial value. The policy assigns one
lever to each error term of Eq.~\eqref{eq:eloc}. Timestep adaptation controls the splitting error,
degree adaptation controls the truncation error and norm defect, and shot
allocation controls the stochastic phase error.

\section{NLSE operating regime}
\label{sec:nlse}

Full-field intensity access at controlled precision is essential in strongly
nonlinear regimes, in which truncated linear embeddings lose convergence
\cite{liu2021,krovi2023}. The measure--update--reload cycle of
Fig.~\ref{fig:hybrid} provides this access, thereby enabling the controller of
Sec.~\ref{sec:cost} to distribute a fixed measurement budget across timestep,
degree, and shots. We benchmark in the strongly focusing regime, since benign
operating ranges do not discriminate between a fragile and a robust nonlinear
approximation, as confirmed by the sweeps of Sec.~\ref{app:benign}. We note that
the nonadaptive solver provides the same hybrid pipeline at a fixed timestep and at
the lowest polynomial degree of Sec.~\ref{sec:cost}, with the controller disabled.
However, it is not an external algorithm, so the shot accounting of
Sec.~\ref{app:diagnostics} isolates the effect of adaptation alone.

The controller response follows from Fig.~\ref{fig:accuracy}. The controller
maintains the local
phase $\theta$ below the truncation and norm-defect thresholds by reducing the
timestep and raising the degree precisely where self-focusing amplifies the
intensity, a response that is unavailable at fixed timestep and degree. The
rejection feedback thereby reallocates the same measurements toward smaller
timesteps and higher degrees during the focusing window, where each shot carries
the greatest phase information, consistent with the budget-allocation argument of
Sec.~\ref{sec:intro}.

The coherent kernels realizing this cycle are validated independently on IBM
Heron-architecture hardware in Sec.~\ref{app:hardware}. The shot--precision
scaling and the conserved-quantity diagnostics are deferred to
Sec.~\ref{app:diagnostics}.

\section{Extension to the Burgers' equation}
\label{sec:burgers}

The viscous Burgers' equation \cite{burgers1974},
\begin{equation}
u_t + u\,u_x = \nu\,u_{xx},
\label{eq:burgers}
\end{equation}
in which subscripts denote partial derivatives, $u(x,t)$ is the velocity field and
$\nu$ the viscosity, serves as a structurally distinct test of the measurement
bottleneck. In two dimensions the field carries components $(u,v)$, each advanced by
the construction below. Three features distinguish this case from the NLSE. The
field is real, the nonlinearity transports rather than dephases, and the linear
step is a non-unitary diffusion. First-order Lie--Trotter splitting is employed;
the splitting order affects only the order of the truncation error and leaves the
measurement, reload, and post-selection cost structure of Sec.~\ref{sec:resource}
unchanged.

For the inviscid transport step, a fluid element moving along its characteristic
$dx/dt=u$ retains a constant value along that path. The system is therefore updated
by back-traced linear interpolation,
\begin{equation}
u_i^{(m+1)} = (1-\alpha_i)\,u_{j_i^0}^{(m)} + \alpha_i\,u_{j_i^1}^{(m)},
\qquad
j_i = i - \frac{u_i^{(m)}\,\Delta t}{\Delta x},
\label{eq:interp}
\end{equation}
in which the superscript $(m)$ indexes the timestep, $j_i$ is the real-valued
departure point traced back from grid node $i$, $j_i^0=\lfloor j_i\rfloor$ and
$j_i^1=(j_i^0+1)\bmod N$ are the two grid nodes bracketing it under periodic
boundaries, and $\alpha_i=j_i-j_i^0$ is the interpolation weight. Rearranging
Eq.~\eqref{eq:interp} as
$u_i^{(m+1)}=u_{j_i^0}^{(m)}+\alpha_i(u_{j_i^1}^{(m)}-u_{j_i^0}^{(m)})$ shows that
one data-dependent product suffices per node, which is evaluated by the EHands
primitive \cite{balewski2025}, while all index arithmetic is performed classically.
In two dimensions the update becomes bilinear interpolation over four neighbors,
applied once per velocity component.

The diffusion step applies the Fourier damping $d_k=e^{-\nu k^2\Delta t}<1$ at
wavenumber $k$, which is non-unitary and therefore cannot be realized by a circuit
alone. Adjoining one ancilla and applying the isometry
$|k\rangle|0\rangle\mapsto|k\rangle\bigl(d_k|0\rangle
+\sqrt{1-d_k^2}\,|1\rangle\bigr)$ restores unitarity, and the damped field is
recovered by retaining the outcome $\mathrm{anc}=0$. For a field normalized as
$\sum_k|\hat{u}(k)|^2=1$, that outcome occurs with probability
\begin{equation}
P(\mathrm{anc}=0) = \sum_k d_k^2\,|\hat{u}(k)|^2,
\label{eq:postselect}
\end{equation}
where $\hat{u}(k)$ denotes the discrete Fourier transform of the field. The
probability approaches unity for small $\nu\Delta t$, so the post-selection is
rarely rejected, and it constitutes the sole resource term absent for the unitary
NLSE dispersion.

\begin{figure}[tbp]
\centering
\includegraphics[height=73.5pt,trim={2.6pt 162.5pt 4.1pt 1.9pt},clip]{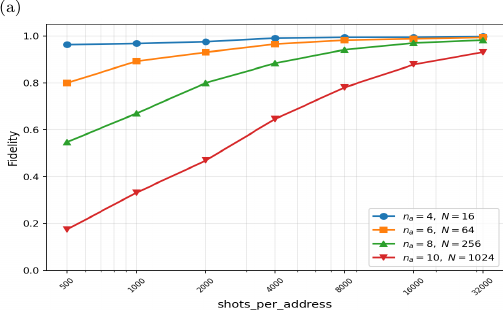}%
\hfill
\includegraphics[height=73.5pt,trim={2.6pt 7.0pt 3.4pt 159.4pt},clip]{fig5}
\caption{One-dimensional Burgers' benchmark with a sinusoidal initial condition at
grid spacing $\Delta x=1$. (a)~Single-step fidelity versus the number of measurement
shots per address, for grid sizes $N\in\{16,64,256,1024\}$ at viscosity $\nu=0.4$.
(b)~Steepening of the sinusoid into a shock precursor at $\nu=0.01$ and
$\Delta t=\Delta x=0.1$, evolved to the inviscid shock-formation time.}
\label{fig:burgers}
\end{figure}

Because the damping contracts the field at every step, the accumulated shot noise
remains bounded. The fidelity decreases over the initial Lie--Trotter steps as
interpolation shot noise accumulates, and subsequently saturates at a plateau
determined by the shot budget, in one and two dimensions alike
(Sec.~\ref{app:diagnostics}). At a fixed budget, the single-step fidelity decreases
with grid size, Fig.~\ref{fig:burgers}(a), which is the empirical signature of the
$\Omega(N/\varepsilon^2)$ readout floor of Sec.~\ref{sec:cost}, while the pipeline
resolves the steepening of the sine wave into a shock precursor,
Fig.~\ref{fig:burgers}(b). The measure--update--reload structure and its resource
signature therefore remain unchanged under variation of the equation, the field
type, and the dimension.

\section{Resource comparison}
\label{sec:resource}

The advantage from a quantum pipeline is determined by the total work of one step.
The full classical and quantum per-step costs are therefore compared at fixed grid
size $N$ and fixed field precision $\varepsilon$, for both the NLSE and the
Burgers' equation.

\emph{a.~Classical cost.} One classical split step on $N=2^n$ points requires
$\mathcal{O}(N\log N)$ floating-point operations, dominated by the two fast Fourier
transforms \cite{cooley1965} of the linear step \cite{taha1984,weideman1986}. The
arithmetic is deterministic at machine precision, and the state occupies
$\mathcal{O}(N)$ memory.

\emph{b.~Quantum cost.} The quantum pipeline encodes the field in
$n+\mathcal{O}(1)$ qubits, an exponential compression in space, and each load or
arithmetic call exhibits CX-circuit depth $\mathcal{O}(N)$ \cite{balewski2024}. The
decisive cost is sampling. Throughout, work counts elementary operations, namely
gate executions for the quantum pipeline and floating-point operations for the
classical solver. Recovery of the full intensity profile to precision
$\varepsilon$ requires $\Omega(N/\varepsilon^2)$ shots per step by the floor
established in Sec.~\ref{sec:cost}, and each shot re-executes the
$\mathcal{O}(N)$-depth preparation, so the total quantum work per step is
\begin{equation}
W_Q = \underbrace{\mathcal{O}(N)}_{\text{depth}}
\times \underbrace{\mathcal{O}(N/\varepsilon^2)}_{\text{shots}}
= \mathcal{O}\!\left(\frac{N^2}{\varepsilon^2}\right),
\label{eq:wq}
\end{equation}
against the classical per-step work $W_{\rm cl}=\mathcal{O}(N\log N)$. Their ratio
is
\begin{equation}
\frac{W_Q}{W_{\rm cl}} = \mathcal{O}\!\left(\frac{N}{\varepsilon^2\log N}\right),
\label{eq:ratio}
\end{equation}
which grows without bound as $N$ increases.

The logarithmic qubit count is therefore the sole asymptotic advantage of the
pipeline. However, we note that the coherent linear step retains a genuine
advantage in isolation, with QFT depth $\mathcal{O}((\log N)^2)$ against the
classical $\mathcal{O}(N\log N)$. Specifically, that advantage is nullified
end to end because the $\mathcal{O}(N)$-depth reloads flanking the step and the
$\Omega(N/\varepsilon^2)$ full-field readout both dominate it. The direct
split-step solver thus affords no end-to-end advantage over its classical
counterpart at fixed precision, for both equations and independently of hardware
quality. Table~\ref{tab:resource} summarizes the comparison, in which the Burgers'
diffusion step additionally incurs the post-selection cost of
Eq.~\eqref{eq:postselect}, absent for the unitary NLSE dispersion.

\begin{table}[tbp]
\caption{Per-step resource comparison for the direct split-step route at grid size
$N=2^n$ and field precision $\varepsilon$. Classical entries count floating-point
operations and the quantum entries count gate executions, except for space, which
counts words and qubits respectively. ``0'' denotes exact and deterministic.}
\label{tab:resource}
\begin{ruledtabular}
\begin{tabular}{lcc}
Resource & Classical & Quantum \\
\colrule
Space            & $\mathcal{O}(N)$ words  & $\mathcal{O}(\log N)$ qubits \\
Linear step      & $\mathcal{O}(N\log N)$   & $\mathcal{O}((\log N)^2)$ depth \\
Reload           & n/a                      & $\mathcal{O}(N)$ depth \\
Shots per step   & $0$                      & $\Omega(N/\varepsilon^2)$ \\
Total work/step  & $\mathcal{O}(N\log N)$    & $\mathcal{O}(N^2/\varepsilon^2)$ \\
Stochastic noise & $0$                      & $\sim 1/\sqrt{S}$ per site \\
\end{tabular}
\end{ruledtabular}
\end{table}

\section{Conclusion}
\label{sec:conclusion}

We present a direct split-step solver that renders the measurement and reload costs
of quantum nonlinear-wave simulation explicit and controllable as budget levers of
a single latency-and-error model. The formulation allows a fixed measurement budget
to be allocated adaptively across timestep, polynomial degree, and shots, and we
validate its coherent kernels on superconducting hardware
(Sec.~\ref{app:hardware}). We find the same measurement bottleneck in both the
NLSE and the Burgers' equations, and we therefore identify a coherent,
measurement-free nonlinear kick as a quantitative target for end-to-end quantum
advantage.

\begin{acknowledgments}
The authors acknowledge discussions with collaborators on quantum data loading,
polynomial arithmetic, and hybrid PDE solvers.
\end{acknowledgments}

\paragraph*{Author contributions.}
Z.G. conceived the project, developed the algorithmic framework and implemented
the simulator benchmark for NLSE. V.D. developed the framework for Burgers'
equation, implemented and benchmarked the results and performed complexity
analysis. A.K., A.C., and R.L. suggested the hardware experiment and sponsored
access to the quantum processor. Z.P. helped refine the idea and revised the
manuscript. All authors discussed the results and approved the final manuscript.

\paragraph*{Data availability.}
The simulator scripts, benchmark data, and figure-generation workflow are described
in the Supplemental Material. The code, benchmark data, and figure scripts are
openly available on Zenodo at \href{https://doi.org/10.5281/zenodo.21245905}%
{doi:10.5281/zenodo.21245905} which resolves to the latest version of the archive.

\paragraph*{Competing interests.}
The authors declare no competing interests.

\appendix

\section{Nonlinear kick versus a QSVT baseline}
\label{app:qsvt}

\begin{figure}[!ht]
\centering
\includegraphics[width=\columnwidth]{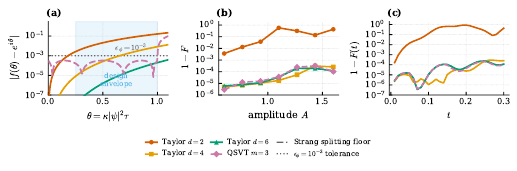}
\caption{Taylor kick versus a QSVT/Chebyshev block encoding of the same phase
oracle $e^{i\kappa|\psi|^2\tau}$, both at phase tolerance $\epsilon_\phi=10^{-3}$
over the design envelope $\Theta\in[0.25,1]$, for $\kappa=150$,
$\Delta t=10^{-2}$, final time $T=0.3$, soliton width $0.75$, and a
$\Delta t=5\times10^{-4}$ reference. (a)~Per-step phase-factor error versus the
local phase $\theta$; the dotted line marks $\epsilon_\phi$. (b)~Final infidelity
$1-F$ versus initial amplitude $A$. (c)~Running infidelity at $A=1.6$; the
dash-dot curve is the Strang splitting floor.}
\label{fig:qsvt}
\end{figure}

The accuracy of the polynomial kick of Eq.~\eqref{eq:polykick} is determined by
its truncation tail alone, provided that competing implementations are held to a
common tolerance. A QSVT \cite{gilyen2019} Chebyshev block encoding of the same
diagonal phase oracle $\exp(i\kappa|\psi|^2\tau)$ is considered, held to the common
uniform phase tolerance $\epsilon_\phi=10^{-3}$ over the design envelope
$\Theta=|\kappa|s_{\max}\tau\in[0.25,1]$. The admissible degree follows from the
Jacobi--Anger expansion, which fixes the smallest Chebyshev quantum signal
processing (QSP) degree \cite{low2017} for which the truncation tail meets a given
tolerance. The block-encoding normalization $s_{\max}$ denotes the peak intensity
attained along the evolution and is set to that peak, so that the polynomial is
evaluated only within the design interval $\Theta\in[0.25,1]$. The admissible QSVT
degree at $\epsilon_\phi=10^{-3}$ is then $m=3$. At matched tolerance, the degree-6
Taylor kick and the degree-3 QSVT polynomial are indistinguishable in fidelity,
both saturating the Strang splitting floor, as shown in Fig.~\ref{fig:qsvt}.
Accuracy therefore does not distinguish the two implementations, and the comparison
of Sec.~\ref{sec:resource} rests on resource cost alone.

\section{Splitting and dispersion ablations}
\label{app:ablation}

\begin{figure}[tbp]
\centering
\includegraphics[width=\columnwidth]{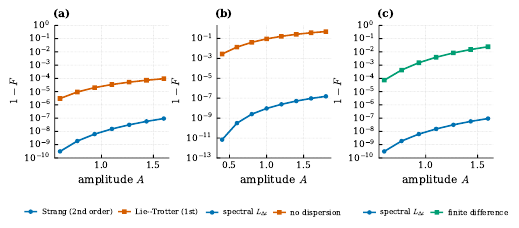}
\caption{Backbone ablations at $\kappa=4$, $\Delta t=0.02$, $T=0.6$, soliton width
$0.75$, against a $\Delta t=0.002$ reference. Final infidelity $1-F$ versus initial
amplitude $A$ for (a)~Strang versus Lie--Trotter splitting, (b)~spectral dispersion
versus no dispersion, and (c)~spectral versus finite-difference dispersion.}
\label{fig:backbone}
\end{figure}

\begin{figure}[tbp]
\centering
\includegraphics[width=\columnwidth]{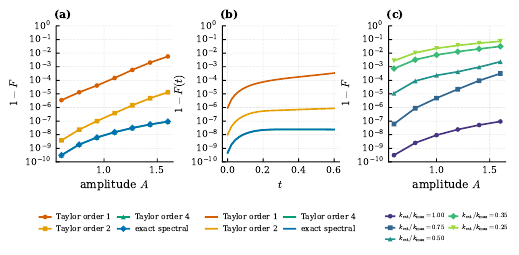}
\caption{Linear-step ablations at $\kappa=4$, $\Delta t=0.02$, $T=0.6$.
(a)~Final infidelity for linear-Taylor propagators of order $1$, $2$, $4$ versus
the exact spectral step. (b)~Running infidelity at $A=1.2$. (c)~Spectral low-pass
filtering at cut fractions $k_{\rm cut}/k_{\max}\in[0.25,1]$.}
\label{fig:lintaylor}
\end{figure}

This appendix establishes that the linear backbone of Eq.~\eqref{eq:splitstep},
held fixed throughout the main text, is not the failure mode. In these ablations
the backbone constitutes the component under test, while the nonlinear kick is
held exact. Three variations are considered: first-order splitting, removal of the
spectral dispersion step, and a finite-difference surrogate for dispersion. Upon
removal of the dispersion step, the evolution degrades progressively with
amplitude, and the finite-difference surrogate trails the spectral implementation
at strong amplitudes. Strang splitting with spectral dispersion is therefore
retained throughout, as shown in Fig.~\ref{fig:backbone}. In a benign regime, a
linear-Taylor surrogate and a spectral low-pass filter both produce negligible
deviations, as shown in Fig.~\ref{fig:lintaylor}, so that the dominant instability
originates in the nonlinear truncation--reload loop rather than in the fine
structure of the linear propagator.

\section{Benign-regime sweeps}
\label{app:benign}

\begin{figure}[tbp]
\centering
\includegraphics[width=\columnwidth]{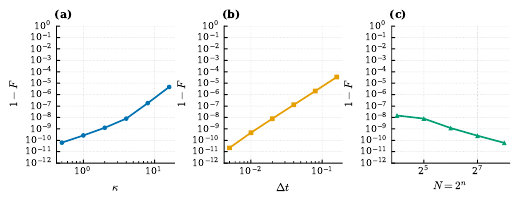}
\caption{Final infidelity $1-F$ of the fixed degree-2 solver at soliton amplitude
$A=1$ and width $0.75$. (a)~Nonlinearity sweep $\kappa\in[0.5,16]$ at
$\Delta t=0.02$. (b)~Timestep sweep $\Delta t\in[0.005,0.16]$ at $\kappa=4$,
$T=0.48$. (c)~Grid-size sweep $N=2^4$--$2^8$ at $\kappa=4$, $\Delta t=0.02$.}
\label{fig:benign}
\end{figure}

Benign operating ranges motivate the strongly focusing benchmark of
Sec.~\ref{sec:nlse} through their saturation. The fixed degree-2 solver is accurate
provided that the nonlinear phase per step, $\theta=\kappa|\psi_j|^2\Delta t$ of
Eq.~\eqref{eq:polykick}, remains small compared with unity, a condition satisfied
throughout the ranges swept here at fixed amplitude $A=1$. These ranges comprise
nonlinearity strength $\kappa\in[0.5,16]$, timestep $\Delta t\in[0.005,0.16]$, and
grid resolution $N\in[16,256]$, as shown in Fig.~\ref{fig:benign}. The final
infidelity remains below $10^{-4}$ across these ranges. Benign regimes are
consequently unable to distinguish a fragile nonlinear approximation from a robust
one. The same saturation is observed for Gaussian, chirped-Gaussian, and
random-phase initial conditions, and for single- as against double-precision
arithmetic, all of which are reproducible from the software artifact.

\section{Hardware validation of the boundary kernels}
\label{app:hardware}

The coherent boundary kernels are assessed on present hardware through
implementation of the intensity-sampling kernel (\textbf{StateReload} together with
readout) and the dispersion kernel (\textbf{SpecProp}) on the 156-qubit
Heron-architecture \texttt{ibm\_pittsburgh} backend at $n=5$, $N=32$, $A=1.8$,
$\Delta t=0.16$, with $1024$ shots per circuit. The remaining nonlinear update
constitutes a classical stage of the measure--update--reload cycle, so that no
modification of the algorithm is required. The experiment was conducted using the
real-time gate and readout error rates and the CLOPS throughput metric that
\texttt{ibm\_pittsburgh} reports at job submission \cite{wack2021}, rather than
published specifications, since superconducting-qubit calibration drifts between
backend accesses. Circuits are transpiled via Qiskit's preset pass manager at
optimization level 1 with fixed seed and executed through the
\texttt{qiskit-ibm-runtime} SamplerV2 primitive in a single job. Agreement with the
ideal statevector distribution is quantified by the classical (Bhattacharyya)
fidelity $F_{\rm cl}=\bigl(\sum_j\sqrt{\widehat{p}_j\,p_j^{\rm ideal}}\bigr)^2$ and
the total-variation distance $D_{\rm TV}$ in Table~\ref{tab:hardware}. The shallow
prepare-and-measure kernel ($65$ two-qubit gates after transpilation) attains
$F_{\rm cl}=0.89$, whereas the deeper dispersion kernel ($282$ two-qubit gates)
attains $0.59$, which quantifies the gate-level cost of the coherent spectral step.
The larger infidelity of the dispersion kernel is consistent with crosstalk error,
which increases with the number of two-qubit gates driven concurrently on
overlapping qubit neighborhoods during the quantum Fourier transform, and is
therefore not attributable to the higher two-qubit gate count alone. Error
mitigation such as measurement twirling or dynamical decoupling was not enabled at
submission, and the controller diagnostics of Sec.~\ref{app:diagnostics} remain
simulator-based.

\begin{table}[tbp]
\caption{Measured agreement of the two boundary kernels with their ideal
distributions on \texttt{ibm\_pittsburgh} ($1024$ shots).}
\label{tab:hardware}
\begin{ruledtabular}
\begin{tabular}{lccc}
Kernel & Two-qubit gates & $F_{\rm cl}$ & $D_{\rm TV}$ \\
\colrule
Intensity sampling  & $65$  & $0.89$ & $0.12$ \\
Spectral dispersion & $282$ & $0.59$ & $0.43$ \\
\end{tabular}
\end{ruledtabular}
\end{table}

\section{Additional benchmark diagnostics}
\label{app:diagnostics}

\begin{figure}[tbp]
\centering
\includegraphics[width=\columnwidth]{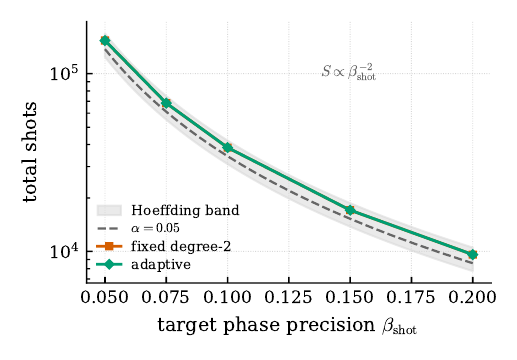}
\caption{Accumulated measurement shots versus target phase precision
$\beta_{\rm shot}$ at $A=1.8$, $N=32$, $\kappa=50$. The band is the Hoeffding bound
of Eq.~\eqref{eq:hoeffding} for failure probabilities $\alpha\in[0.01,0.10]$; the
dashed curve marks $\alpha=0.05$.}
\label{fig:shots}
\end{figure}

\begin{figure}[tbp]
\centering
\includegraphics[width=\columnwidth]{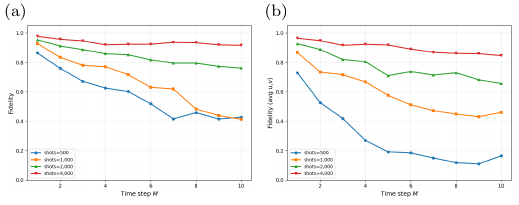}
\caption{Burgers' error accumulation over $M=10$ Lie--Trotter steps at
$\Delta t=0.1$, $\Delta x=1$, $\nu=0.4$, sine initial condition, for shot budgets
$S\in\{500,1000,2000,4000\}$ per address. (a)~1D at $N=64$. (b)~2D on an
$8\times8$ grid, fidelity averaged over the velocity components $(u,v)$.}
\label{fig:burgersacc}
\end{figure}

The sampling cost of the benchmark of Sec.~\ref{sec:nlse} is determined by the
target phase precision alone. The accumulated shot count follows the
$\beta_{\rm shot}^{-2}$ scaling of Eq.~\eqref{eq:phaseerr} as the target is
tightened, as shown in Fig.~\ref{fig:shots} together with the Hoeffding bound of
Eq.~\eqref{eq:hoeffding}. The close agreement between the adaptive and nonadaptive
curves confirms that adaptation does not remove the sampling bottleneck. In the stored run, the controller records
one rejected attempt, one shrink event, and two growth events at $A=1.8$. A
complementary diagnostic for future hardware studies is the drift of the cubic
NLSE invariants, namely the mass $M=\sum_j|\psi_j|^2\Delta x$, the momentum
$P=\mathrm{Im}\sum_j\bar{\psi}_j(\partial_x\psi)_j\Delta x$, and the energy
$E=\sum_j\bigl[\tfrac{1}{2}|(\partial_x\psi)_j|^2-\tfrac{\kappa}{2}|\psi_j|^4\bigr]
\Delta x$. The truncated kick followed by renormalization biases these quantities
even when the fidelity remains high. The norm-defect threshold $\delta$ constrains
the mass drift directly, whereas $P$ and $E$ are not constrained by the present
formulation. For the multistep Burgers' evolution of Sec.~\ref{sec:burgers},
viscous contraction bounds the accumulated error. The fidelity loss induced by
interpolation shot noise saturates at a value determined by the shot budget,
independently of dimension, as shown in Fig.~\ref{fig:burgersacc}.


\begin{thebibliography}{99}

\bibitem{strang1968}
G.~Strang, \emph{On the construction and comparison of difference schemes},
SIAM J. Numer. Anal. \textbf{5}, 506 (1968).

\bibitem{lloyd1996}
S.~Lloyd, \emph{Universal quantum simulators}, Science \textbf{273}, 1073 (1996).

\bibitem{harrow2009}
A.~W. Harrow, A.~Hassidim, and S.~Lloyd, \emph{Quantum algorithm for linear systems
of equations}, Phys. Rev. Lett. \textbf{103}, 150502 (2009).

\bibitem{lloyd2020}
S.~Lloyd, G.~D. Palma, C.~Gokler, B.~Kiani, Z.-W. Liu, M.~Marvian, F.~Tennie, and
T.~Palmer, \emph{Quantum algorithm for nonlinear differential equations}, arXiv
preprint arXiv:2011.06571 (2020).

\bibitem{liu2021}
J.-P. Liu, H.~{\O}. Kolden, H.~K. Krovi, N.~F. Loureiro, K.~Trivisa, and A.~M.
Childs, \emph{Efficient quantum algorithm for dissipative nonlinear differential
equations}, Proc. Natl. Acad. Sci. USA \textbf{118}, e2026805118 (2021).

\bibitem{jin2024bull}
S.~Jin and N.~Liu, \emph{Quantum algorithms for nonlinear partial differential
equations}, Bull. Sci. Math. \textbf{194}, 103457 (2024).

\bibitem{agrawal2019}
G.~P. Agrawal, \emph{Nonlinear Fiber Optics}, 6th ed. (Academic Press, San Diego,
2019).

\bibitem{pitaevskii2003}
L.~P. Pitaevskii and S.~Stringari, \emph{Bose--Einstein Condensation} (Oxford
University Press, Oxford, 2003).

\bibitem{burgers1974}
J.~M. Burgers, \emph{The Nonlinear Diffusion Equation: Asymptotic Solutions and
Statistical Problems} (D.~Reidel, Dordrecht, 1974).

\bibitem{leyton2008}
S.~K. Leyton and T.~J. Osborne, \emph{A quantum algorithm to solve nonlinear
differential equations}, arXiv preprint arXiv:0812.4423 (2008).

\bibitem{krovi2023}
H.~Krovi, \emph{Improved quantum algorithms for linear and nonlinear differential
equations}, Quantum \textbf{7}, 913 (2023).

\bibitem{wu2025}
H.-C. Wu, J.~Wang, and X.~Li, \emph{Quantum algorithms for nonlinear dynamics:
Revisiting Carleman linearization with no dissipative conditions}, SIAM J. Sci.
Comput. \textbf{47}, A943 (2025).

\bibitem{budinski2021}
L.~Budinski, \emph{Quantum algorithm for the advection--diffusion equation
simulated with the lattice Boltzmann method}, Quantum Inf. Process. \textbf{20}, 57
(2021).

\bibitem{sanavio2024}
C.~Sanavio and S.~Succi, \emph{Lattice Boltzmann--Carleman quantum algorithm and
circuit for fluid flows at moderate Reynolds number}, AVS Quantum Sci. \textbf{6},
023802 (2024).

\bibitem{joseph2020}
I.~Joseph, \emph{Koopman--von Neumann approach to quantum simulation of nonlinear
classical dynamics}, Phys. Rev. Research \textbf{2}, 043102 (2020).

\bibitem{jin2024prl}
S.~Jin, N.~Liu, and Y.~Yu, \emph{Quantum simulation of partial differential
equations via Schr\"odingerization}, Phys. Rev. Lett. \textbf{133}, 230602 (2024).

\bibitem{jin2023pra}
S.~Jin, N.~Liu, and Y.~Yu, \emph{Quantum simulation of partial differential
equations: Applications and detailed analysis}, Phys. Rev. A \textbf{108}, 032603
(2023).

\bibitem{jin2026}
S.~Jin, N.~Liu, and C.~Ma, \emph{Hybrid quantum-classical algorithms for complex
nonlinear partial differential equations with Ginzburg--Landau potential and vortex
motion laws}, arXiv preprint arXiv:2604.14079 (2026).

\bibitem{lubasch2020}
M.~Lubasch, J.~Joo, P.~Moinier, M.~Kiffner, and D.~Jaksch, \emph{Variational
quantum algorithms for nonlinear problems}, Phys. Rev. A \textbf{101}, 010301
(2020).

\bibitem{taha1984}
T.~R. Taha and M.~J. Ablowitz, \emph{Analytical and numerical aspects of certain
nonlinear evolution equations. II. Numerical, nonlinear Schr\"odinger equation},
J. Comput. Phys. \textbf{55}, 203 (1984).

\bibitem{guo2025}
Z.~Guo, Z.~Pan, A.~Khan, and J.~Balewski, \emph{Vectorized attention with
learnable encoding for quantum transformer}, arXiv preprint arXiv:2508.18464
(2025).

\bibitem{grover2002}
L.~Grover and T.~Rudolph, \emph{Creating superpositions that correspond to
efficiently integrable probability distributions}, arXiv preprint
arXiv:quant-ph/0208112 (2002).

\bibitem{araujo2021}
I.~F. Araujo, D.~K. Park, F.~Petruccione, and A.~J. da Silva, \emph{A
divide-and-conquer algorithm for quantum state preparation}, Sci. Rep. \textbf{11},
6329 (2021).

\bibitem{balewski2024}
J.~Balewski, M.~G. Amankwah, R.~Van Beeumen, E.~W. Bethel, T.~Perciano, and
D.~Camps, \emph{Quantum-parallel vectorized data encodings and computations on
trapped-ion and transmon QPUs}, Sci. Rep. \textbf{14}, 3435 (2024).

\bibitem{javadi2024}
A.~Javadi-Abhari, M.~Treinish, K.~Krsulich, C.~J. Wood, J.~Lishman, J.~Gacon,
S.~Martiel, P.~D. Nation, L.~S. Bishop, A.~W. Cross, B.~R. Johnson, and J.~M.
Gambetta, \emph{Quantum computing with Qiskit}, arXiv preprint arXiv:2405.08810
(2024).

\bibitem{coppersmith1994}
D.~Coppersmith, \emph{An Approximate Fourier Transform Useful in Quantum
Factoring}, Tech. Rep. RC19642 (IBM Research Division, Yorktown Heights, NY, 1994).

\bibitem{balewski2025}
J.~Balewski, C.~Pestano, M.~G. Amankwah, E.~W. Bethel, T.~Perciano, and R.~Van
Beeumen, \emph{EHands: Quantum protocol for polynomial computation on real-valued
encoded states}, arXiv preprint arXiv:2502.15928 (2025).

\bibitem{weng2026}
K.~Weng, Z.~Meng, and G.~Hu, \emph{Quantum computing of the nonlinear
Schr\"odinger equation via measurement-induced potential reconstruction}, arXiv
preprint arXiv:2601.19184 (2026).

\bibitem{canonne2020}
C.~L. Canonne, \emph{A short note on learning discrete distributions}, arXiv
preprint arXiv:2002.11457 (2020).

\bibitem{gilyen2019}
A.~Gily\'en, Y.~Su, G.~H. Low, and N.~Wiebe, \emph{Quantum singular value
transformation and beyond: exponential improvements for quantum matrix
arithmetics}, in \emph{Proceedings of the 51st Annual ACM SIGACT Symposium on
Theory of Computing} (ACM, New York, 2019), p.~193.

\bibitem{low2017}
G.~H. Low and I.~L. Chuang, \emph{Optimal Hamiltonian simulation by quantum signal
processing}, Phys. Rev. Lett. \textbf{118}, 010501 (2017).

\bibitem{cooley1965}
J.~W. Cooley and J.~W. Tukey, \emph{An algorithm for the machine calculation of
complex Fourier series}, Math. Comput. \textbf{19}, 297 (1965).

\bibitem{weideman1986}
J.~A.~C. Weideman and B.~M. Herbst, \emph{Split-step methods for the solution of
the nonlinear Schr\"odinger equation}, SIAM J. Numer. Anal. \textbf{23}, 485
(1986).

\bibitem{wack2021}
A.~Wack, H.~Paik, A.~Javadi-Abhari, P.~Jurcevic, I.~Faro, J.~M. Gambetta, and
B.~R. Johnson, \emph{Quality, speed, and scale: three key attributes to measure the
performance of near-term quantum computers}, arXiv preprint arXiv:2110.14108
(2021).

\end{thebibliography}
\end{document}